\documentclass[aps,prd,nofootinbib,preprint]{revtex4}
\usepackage[colorlinks=true, pdfstartview=FitV, linkcolor=blue, citecolor=red, urlcolor=magenta]{hyperref}
\usepackage{graphicx}
\usepackage{latexsym}
\usepackage{amsmath}
\usepackage{amsfonts}
\usepackage{amssymb}
\usepackage{verbatim}
\usepackage{dcolumn}
\usepackage{amssymb}
\usepackage{bm}
\usepackage[english]{babel}
\newcommand{\be}{\begin{equation}}
\newcommand{\ee}{\end{equation}}
\newcommand{\bea}{\begin{eqnarray}}
\newcommand{\eea}{\end{eqnarray}}

\begin{document}

\newcommand{\JPess}{
\affiliation{Departamento de F\'isica, Universidade Federal da Para\'iba, \\Caixa Postal 5008, 58059-900, Jo\~ao Pessoa, PB, Brazil}
}

\title{Induced Brownian motion by the  Friedmann-Robertson-Walker spacetime in the presence of a cosmic string}

\author{Herondy F. Santana Mota}
\email{hmota@fisica.ufpb.br}
\JPess
\author{Eug\^enio R. Bezerra de Mello}
\email{emello@fisica.ufpb.br}
\JPess

\begin{abstract}
In this paper we investigate the quantum Brownian motion of a massive scalar point particle induced by the FRW spacetime in the presence of a linear topological defect named cosmic string. In addition, we also consider a flat boundary orthogonal to the defect to analyse its effect on the particle's motion. For both cases we found exact expressions for the renormalized mean square deviation of the particle velocity, the quantity that measures the induced Brownian motion, and obtain asymptotic expressions when the point particle is near and far away from the cosmic string and the boundary. Furthermore, in both cases, there appears divergencies in the mean square deviation of the particle velocity coming from a time interval that correspond to a round trip of the massive point particle between its position and the cosmic string/flat boundary. The number of divergencies depends upon the radial position of the particle and the parameter associated with the cosmic string in the case without boundary, and upon this parameter and the radial and $z$ positions of the particle in the case with boundary. 
\end{abstract}

%\pacs{11.30.Cp, 11.30.Qc, 03.65.Pm}

\maketitle
%%%%%%%%%%%%%%%%%
\section{Introduction}
%%%%%%%%%%%%%%%%%%%
In the context of quantum field theory, the existence of quantum vacuum fluctuations of relativistic fields gives rises to a variety of phenomena. In fact, it is well known that the quantum vacuum state plays a fundamental role, not only in a microscopic level, but also in a macroscopic one. One of the very known and already observed effect, at least for the electromagnetic field, is the Casimir effect that comes about due to modifications in the fluctuations of the quantum vacuum of the field once boundary conditions are implemented. These modifications also occur if the field propagates on a curved background and, thus, a gravitational analog of the Casimir effect also takes place \cite{Mostepanenko:1997sw, bordag2009advances, Milton:2001yy}. One can also mention, as alternative examples, fluctuations on the classical light-cone once the spacetime is quantized \cite{Ford:1994cr, Mota:2016mhe} and induced quantum motions of charged particles \cite{Camargo:2017scp, Gour:1998my, Ford:2005rs, Bessa:2008pr, Yu:2004gs, Yu:2004tu, DeLorenci:2014zna}. Moreover, the analysis of self-interactions on a charged point-particle in curved spacetimes has been considered in Refs. \cite{Burko:1999zy,Wiseman:2000rm, BezerradeMello:2012nq}.

It is well known that the phenomenon of random motion of suspended particles  in a fluid, i.e., the Brownian motion, happens as a consequence of thermal fluctuations that characterize the unceasing motion of the molecules that constitute the fluid \cite{Reichl}. In quantum field theory, nevertheless, the relativistic fields possess a nonzero energy even at zero temperature (their ground state). Thereby, the same quantum fluctuations that lead the fields to have a ground state energy, will make a particle interacting with the field to undergo random motion with a nonzero dispersion velocity \cite{Gour:1998my, Ford:2005rs}. The quantum Brownian motion caused by fluctuations of the electromagnetic and scalar fields has been investigated in Refs. \cite{Bessa:2008pr, Yu:2004gs, Yu:2004tu} and Refs. \cite{DeLorenci:2014zna}. In Refs. \cite{Camargo:2017scp, DeLorenci:2014zna}, for instance, the authors have investigated the nonrelativistic quantum Brownian motion on a charged scalar particle induced by the presence of reflecting boundaries in Minkowski spacetime. In special, in Ref. \cite{Bessa:2008pr}, this phenomenon has been considered in the background of the spatially flat Friedman-Robertson-Walker (FRW) spacetime inducing quantum vacuum fluctuations of the electromagnetic field which, in turn, induce Brownian motion on a charged test particle. 

The spatially-flat FRW spacetime is the best fit for the observational data to describe our large-scale homogeneous and isotropic expanding universe \cite{Ade:2015bva}. It represents the Standard Model of Cosmology and it is also successful in predicting and describing cosmological phenomenon such as the Cosmic Microwave Background Radiation and formation of both large scale structure and light elements (hydrogen, deuterium, helium, etc.) \cite{weinberg}. Thus, the number of reasons why one should consider the spatially-flat FRW spacetime to investigate its effect on quantum field theory systems is beyond doubt countless.

In the present paper we intend to investigate the quantum brownian motion of a charged scalar particle induced by quantum vacuum fluctuations of a massless scalar field in the spatially-flat FRW spacetime. We will also consider the presence of a cosmic string and take into consideration a flat boundary orthogonal to the defect. Cosmic strings are formed due to symmetry breaking phase transitions in the early universe. In this scenario, cosmic strings are predicted in the context of extensions of the Standard Model of Particle Physics \cite{hindmarsh, VS} as well as in the context of string theory \cite{Copeland:2011dx, Hindmarsh:2011qj}. They may have astrophysical, cosmological and gravitational implications, leaving prints during their time-life and, therefore, providing ways of detecting them \cite{hindmarsh, VS,Copeland:2011dx, Hindmarsh:2011qj, Mota:2014uka}. 

Moreover, if a cosmic string is found to be static, straight, thin and very long, one can consider an idealized model to describe the spacetime generated by it. This spacetime has conical topology with a planar angle deficit on the two-surface orthogonal to it given by $\Delta\phi = 8\pi G\mu$, where $G$ is the Newton's gravitational constant and $\mu$ the cosmic string linear energy density \cite{hindmarsh, VS}. Thus, we expect the Brownian motion to be induced not only by the cosmic string topology but also by the flat boundary condition.

The paper is organized as follows: In Sec.\ref{sec2} we present the spatially-flat FRW spacetime in the presence of a cosmic string and its conformal symmetry properties. By means of this symmetry it is possible to connect the FRW spacetime in the presence of a cosmic string with the spacetime purely due to a cosmic string. We show how the Klein-Gordon equation and the Wightman function in both spacetimes are related using the conformal symmetry. The dynamics of a massive scalar point particle under the influence of an external scalar field is also revisited. In Sec.\ref{sec3} we obtain exact expressions for the mean square deviation (MSD) of the particle velocity induced only by the geometry and topology of the FRW and cosmic string spacetimes. We also analyse the influence, on the MSD of the particle velocity, of a flat boundary in the FRW spacetime in the presence of a cosmic string. In Sec.\ref{con} we present our conclusions. We also provide an Appendix to develop some additional calculations. Throughout we use natural units $\hbar=c=1$.
%
%%%%%%%%%%%%%%%%%%%%%%%%%%%%%%%%%%%%
\section{Scalar particle dynamics in FRW spacetime with a cosmic string}
\label{sec2}
%%%%%%%%%%%%%%%%%%%%%%%%%%%%%%%%%%%%
In this section we provide the necessary framework to investigate the induced quantum Brownian motion of a charged scalar particle. We present the structure of the geometry of the spacetime under consideration, and construct the positive-energy Wightman function associated with a the massless scalar field, whose quantum vacuum fluctuations induce the quantum motion. Thus, two specific situations will be considered: the case without and with flat boundary. In the latter case, the corresponding Wightman function is obtained by imposing Dirichlet boundary condition on the massless scalar field. The dynamics of a scalar point particle under the action of a scalar field will also be reviewed.
%
%%%%%%%%%%%%%%%%%%%%%%%%%%%%%%%%%%%
\subsection{The Klein-Gordon equation and Wightman function}
\label{sub2.1}
%%%%%%%%%%%%%%%%%%%%%%%%%%%%%%%%%%%%
It is well known that the (3+1)-dimensional spatially-flat FRW spacetime can be conformally related with the Minkowski spacetime. However, in the case we consider, the FRW spacetime is conformally related with the locally-flat cosmic string spacetime. To show this, let us first present the spatially-flat FRW spacetime in the absence of a cosmic string \cite{Bessa:2008pr, weinberg}, i.e.,
\begin{equation}
ds^2 = g_{\mu\nu}dx^{\mu}dx^{\nu} = dt^2 -a^2(t) \left( dx^2 + dy^2 + dz^2\right),
\label{AFRW}
\end{equation}
where $a(t)$ is the scale factor. 

In order to include a cosmic string in the spacetime described by the line element \eqref{AFRW} it is convenient to write the spatial part of it in cylindrical coordenates. By using this procedure, we can then present the FRW spacetime in the presence of a cosmic string as represented by the following line element:
\begin{equation}
ds^2 = g_{\mu\nu}dx^{\mu}dx^{\nu} = a^2(\eta)\left(d\eta^2 - d\rho^2 - \rho^2d\phi^2 - dz^2\right),
\label{effM}
\end{equation}
where $d\eta = a(t)dt$ is the conformal time, $\rho\geqslant 0$, $\phi \in \lbrack 0,\ 2\pi /p]$ and $(t, \ z)\in (-\infty ,\ \infty )$ define the coordinates range and the paremeter $p>1$ characterizes the presence of the cosmic string. This parameter is given in terms of the linear energy density $\mu$ of the cosmic string and the Newton's gravitational constant $G$ as $p= (1-4G\mu)^{-1}$ \cite{hindmarsh, VS}.

The line element \eqref{effM} describes a conformal locally-flat FRW universe in the presence of a cosmic string. In other words, the metric $g_{\mu\nu}$ characterizing the FRW spacetime is conformally related to the metric $\bar{g}_{\mu\nu}(x)$ characterizing the locally-flat cosmic string spacetime, described only by the line element inside the brackets in Eq. \eqref{effM}. This conformal relation is given through the conformal factor $\Omega = a(\eta)$ as  $g_{\mu\nu} = \Omega^2 \bar{g}_{\mu\nu}(x)$. 

The massless scalar filed modes, propagating in the background represented by the line element \eqref{effM}, non-minimally coupled to gravity, obeys the Klein-Gordon (KG) equation 
\begin{equation}
\left[\frac{1}{\sqrt{|g|}}\partial_{\mu}\left(\sqrt{|g|}g^{\mu\nu}\partial_{\nu}\right) +\xi R\right]\Phi = 0,
\label{KG}
\end{equation}
where $g={\rm det}(g_{\mu\nu})$, $R$ is the Ricci scalar, $\xi$ is the curvature coupling. In the case $\xi =\frac{1}{6}$, the theory is conformally invariant in $(3+1)$-dimensional spacetime. If that is so, the Klein-Gordon equation \eqref{KG} in the spacetime described by $g_{\mu\nu}(x)$ in Eq. \eqref{effM} is related to the Klein-Gordon equation in the cosmic string spacetime characterized by the metric $\bar{g}_{\mu\nu}(x)$ \cite{birrell1984quantum}, that is,
\begin{eqnarray}
\left[\frac{1}{\sqrt{|g|}}D_{\mu}\left(\sqrt{|g|}g^{\mu\nu}D_{\nu}\right)+\frac{1}{6} R\right]\Phi(x)=\Omega^{-3}\left[\frac{1}{\sqrt{|\bar{g}|}}\bar{D}_{\mu}\left(\sqrt{|\bar{g}|}\bar{g}^{\mu\nu}\bar{D}_{\nu}\right)+\frac{1}{6} \bar{R}\right]\bar{\Phi}(x),
\label{KGconformal}
\end{eqnarray}
where $\Omega = a(\eta)$ is the conformal factor and
\begin{equation}
\Phi(x) = \Omega^{-1}(x)\bar{\Phi}(x).
\label{confsol}
\end{equation}
Note that in the cosmic string spacetime $\bar{R}=0$ for $\rho>0$. Therefore, once we solved the KG equation, in the cosmic string spacetime, for the massless scalar field $\bar{\Phi}(x)$, we can afterwords use the relation \eqref{confsol} to obtain the solution $\Phi(x)$ in the FRW spacetime.

The normalized solution $\bar{\Phi}(x)$ for the KG equation in the cosmic string spacetime has been obtained in many different contexts (see for instance \cite{Mota:2017slg} and references therein) and is given by
\begin{equation}
\bar{\Phi}_{\sigma}(x) = \left(\frac{p\eta}{8\pi^2\omega}\right)^{\frac{1}{2}} e^{-i\omega t + inp\phi + ik_zz}J_{p|n|}(\kappa\rho),
\label{solution}
\end{equation}
where $\sigma = (\kappa, n, k_z)$ stands for the set of quantum numbers. Moreover, by making use of the solution \eqref{solution}, the positive-energy Wightman function can be obtained through the sum on the quantum numbers $\sigma$ as
\begin{equation}
W(x,x') = \sum_{\sigma}\bar{\Phi}_{\sigma}(\eta,x^j)\bar{\Phi}_{\sigma}(\eta',x'^j).
\label{wff}
\end{equation}
 In fact, in Appendix \ref{A} we present the renormalized Wightman function associated to the solution \eqref{solution} considering only the cosmic string geometry. This function is represented by $W_{\text{r}}^{(+)}(x,x')$. We also derive the renormalized Wightman function considering a flat boundary, given by Eq. \eqref{wftotal}. This will help us with the calculations we perform in the next section. 
 
Furthermore, the conformal relation between both the Wightman function in the FRW spacetime and the Wightman function in the cosmic string spacetime is given by 
\begin{equation}
W^{(\rm FRW)}(x,x') = a^{-1}(\eta)a^{-1}(\eta')W(x,x'),
\label{twopf}
\end{equation}
where $W(x,x')$ may represent the Wightman function only due to the cosmic string spacetime as well as the Wightman function due to the cosmic string spacetime with a flat boundary (see Appendix \ref{A}).  Let us next present the dynamics of a point particle under the influence of a scalar field.
%
%
%
%%%%%%%%%%%%%%%%%%%%%%%%%%%%%%%%%%%%
\subsection{Dynamics of a point particle and induced Brownian motion}
\label{sub2.2}
%%%%%%%%%%%%%%%%%%%%%%%%%%%%%%%%%%%%
%
The equation of motion for a scalar point particle with mass $m$ and charge $q$ under the influence of a massless scalar field $\Phi$ in a arbitrary curved spacetime is given by \cite{Poisson:2011nh, Quinn:2000wa}
\begin{equation}
\frac{D(mu^{\mu})}{d\tau} = qg^{\mu\nu}\nabla_{\nu}\Phi,
\label{eqmov}
\end{equation}
where $u^{\mu}$ is the particle's four velocity, $\tau$ is the proper time and 
\begin{equation}
\frac{DV^{\mu}}{d\tau} = \frac{dV^{\mu}}{d\tau} + \Gamma^{\mu}_{\nu\rho}V^{\nu}V^{\rho},
\label{coderi}
\end{equation}
is the covariant derivative. In general, the point particle interacts with the scalar field $\Phi$ and, consequently, its mass $m$ varies with the proper time $\tau$, indicating the existence of back-reaction effects \cite{Poisson:2011nh, Quinn:2000wa}. Here, we will assume that the particle behaves like a test particle so that its field contribution is sufficiently weak, which allow us to ignore the back-reaction effects. This assumption is plausible if we consider that the particle moves non-relativistically. In this case, the spatial components of Eq. \eqref{eqmov} are the most relevant ones and we can write it as 
\begin{equation}
m\frac{Du^{i}}{d t} = f^{i} + f^{i}_{\text{ext}},
\label{eqmov2}
\end{equation}
where $f^i = qg^{ij}\nabla_{j}\Phi$ and $f^{i}_{\text{ext}}$ is an external force associated with possible classical contributions of the field $\Phi$. In fact, as we wish to study the Brownian motion induced by quantum vacuum fluctuations, the scalar field $\Phi$ has to be taken as an operator acting on Hilbert space, although it can occasionally have classical contributions. We can thus consider the case where the point particle is bounded by assuming 
\begin{equation}
 f^{i}_{\text{ext}} = \frac{m}{r}(-r^2u^{\phi}u^{\phi}, 2u^{\phi}u^{r},0).
\label{extfor}
\end{equation}
The use of this external force as well as the Christoffel symbols 
\begin{eqnarray}
&&\Gamma^{r}_{r t}=\Gamma^{r}_{t r} = \Gamma^{\phi}_{t\phi} = \Gamma^{\phi}_{\phi t} = \Gamma^{z}_{z t} = \Gamma^{z}_{t z} = \frac{1}{a}\frac{da}{dt},\nonumber\\
&&\Gamma^{\phi}_{r\phi} = \Gamma^{\phi}_{\phi r} = \frac{1}{r},\qquad\qquad\qquad \Gamma^{r}_{\phi\phi} = -r,
\label{CS}
\end{eqnarray}
leads us to the following expression for the velocity-velocity correlation function after the integration of Eq. \eqref{eqmov2}:
\begin{equation}
\langle u^i(\eta,x^{j})u^i(\eta,x'^{j})\rangle = \frac{q^2}{m^2a^4(\eta)}\int_0^{\eta}d\eta_1a^3(\eta_1)\int_0^{\eta} d\eta_2a^3(\eta_2)g^{ii}g^{i'i'}\partial_i\partial_{i'}W^{(\rm FRW)}(x,x'),
\label{sec2eq2}
\end{equation}
where we took the initial velocity to be zero $u^i(\eta_0=0)=0$, $dt =a(\eta)d\eta$ and the expectation value is evaluated considering the vacuum state of the field $\Phi(x)$. The renormalized vacuum fluctuation of the velocity squared is then obtained after subtracting the divergent contribution of the velocity correlation function and then taking the coincident limit $x'^i\rightarrow x^i$, as shown below:
\begin{equation}
\langle (u^i)^2 \rangle_{\text{r}}  =  \lim_{x'^i \rightarrow x^i}[\langle u^i(\eta,x^{j})u^i(\eta,x'^{j})\rangle - \langle u^i(\eta,x^{j})u^i(\eta,x'^{j})\rangle_{\text{d}}].
\label{vsquared}
\end{equation}
Note that, in our case, the divergent contribution in the second term on the r.h.s of the above equation corresponds to the velocity-velocity correlation function in the FRW spacetime. At the Wightman function level, the divergent contribution comes from the first term on the r.h.s of $W^{(+)}(x,x')$ in \eqref{A4}. We should also note that the vacuum expectation value of the field $\Phi(x)$ is zero and as the velocity is given in terms of it through Eq. \eqref{eqmov2} it follows that $ \langle u^i \rangle = 0$. Thereby, the mean square deviation (MSD) of the particle velocity is 
\begin{equation}
\langle (\Delta u^i)^2 \rangle =  \langle (u^i)^2 \rangle - \langle u^i \rangle^2 = \langle (u^i)^2 \rangle.
\label{msd}
\end{equation}
Therefore, the induced Brownian motion is a consequence of a nonzero MSD of the particle velocity.
%%%%%%%%%%%%%%%%%%%%%%%%%%
\section{Induced Brownian motion}
\label{sec3}
%%%%%%%%%%%%%%%%%%%%%%%%%%
%
Since we know the two-point function (Wightman function) of the massless scalar field in the cosmic string spacetime, we can make use of the conformal symmetry to re-write the velocity correlation function in a more workable form. Thus, by substituting Eq. \eqref{twopf} into Eq. \eqref{sec2eq2}  we get 
\begin{equation}
\langle u^i(\eta,x^{j})u^i(\eta,x'^{j})\rangle = \frac{q^2}{m^2a^4(\eta)}\int_0^{\eta}d\eta_1a^2(\eta_1)\int_0^{\eta}d\eta_2a^2(\eta_2)g^{ii}g^{i'i'}\partial_i\partial_{i'}W(x,x').
\label{sec2eq3}
\end{equation}
This expression will allow us to analyse two situations in which quantum Brownian motion will be induced. In one of them, the Brownian motion is caused by quantum vacuum fluctuations due to the conic geometry of the cosmic string while the other will be caused by quantum vacuum fluctuations due to both the cosmic string geometry and a flat boundary. Let us them consider the first case.
%
%%%%%%%%%%%%%%%%%%
\subsection{No boundary}
\label{sec3.1}
%%%%%%%%%%%%%%%%%%
%
We are now in position to use Eq. \eqref{sec2eq3} along with the Wightman function $W^{(+)}(x,x')$ in \eqref{A4} to obtain each component of the MSD of the particle velocity \eqref{msd} in the spacetime time described by the line element \eqref{effM}. Here, the calculation will be considered only taking into consideration the FRW spacetime in the presence of a cosmic string. Upon starting by the radial component, $\rho$, we found that after some intermediate steps its corresponding renormalized expression is given by
\begin{eqnarray}
\langle (\Delta u^{\rho})^2\rangle_{\text{r}} &=& \frac{q^2}{\pi^2m^2\eta^2a^4}\left[\sideset{}{'}\sum_{n=1}^{[p/2]}f^{(1)}_n(R) -2\int_0^{\infty}d\xi g(\xi,p)f^{(1)}_{\xi}(R) \right],
\label{radialC}
\end{eqnarray}
where $[p/2]$ represents the integer part of $p/2$, and the prime on the sign of summation means that in the case $p$ is an integer number the term $n=p/2$ should be taken with the coefficient $1/2$, i.e.,
\begin{eqnarray}
\sideset{}{'}\sum_{n=1}^{[p/2]} \rightarrow \frac{1}{2}\sum_{n=1}^{p-1}.
\label{integer}
\end{eqnarray}
The functions $g(\xi,p)$ and $f^{(1)}_{\gamma}(R)$ are given by
\begin{equation}
g(\xi,p) = \frac{p\sin(p\pi)}{4\pi[\cosh(p\xi) - \cos(p\pi)]},
\label{gfun}
\end{equation}
and
\begin{equation}
f^{(1)}_{\gamma}(R) = -\frac{1}{R^2(1-R^2s_{\gamma}^2)} + \frac{(1 + s_{\gamma}^2)}{4R^3s_{\gamma}^3}\ln\left(\frac{Rs_{\gamma}+1}{Rs_{\gamma}-1}\right)^2.
\label{fun1}
\end{equation}
Note that in our notation $\gamma$ stands for $n$ in the first term in \eqref{radialC} and for $\xi$ in the second term. In addtion, $s_n = \sin(n\pi/p)$, $s_{\xi} = \cosh(\xi/2)$ and $R=\frac{2\rho}{\eta}$. We can also note that for integer values of $p$, the second term on the r.h.s of \eqref{radialC} is absent while for values of $p<2$ the first term is absent. Evidently, if there is no cosmic string spacetime, that is, $p=1$, the renormalized MSD of the particle velocity \eqref{radialC} is zero.

Let us now analyse the limiting cases $R\ll 1$ and $R\gg 1$. In the latter the function $f^{(1)}_{\gamma}$ behaves as
\begin{equation}
f^{(1)}_{\gamma}(R) \simeq \frac{1}{R^4}\left[\frac{1}{s_{\gamma}^4}+\frac{2}{s_{\gamma}^2}\right],
\label{lim1}
\end{equation}
which is an approximated expression valid for the first term on the r.h.s. of Eq. \eqref{radialC} as well as for the second term.

On the other hand, in the limiting case $R\ll 1$, for the first term in the r.h.s \eqref{radialC}, we have
\begin{equation}
f^{(1)}_{n}(R) \simeq \frac{1}{R^2s_{n}^2}.
\label{lim2.1}
\end{equation}
There is no problem to get this behaviour, in this limit, for this term since $s_n$ can be at most one. However, we have to be a bit careful when considering this limit ($R\ll 1$) for the second term on the r.h.s of \eqref{radialC} since  $s_{\xi}$ can grow up to infinity and, thus, there is no guaranty the product $Rs_{\xi}$ will be small in \eqref{fun1}. In order to circumvent this problem, in the limiting process we can consider only integer values of $p$. Thus, we have only the first term on the r.h.s of Eq. \eqref{radialC}, with the approximation given by \eqref{lim2.1}. In this case the leading contribution to the renormalized MSD of the particle velocity is
\begin{eqnarray}
\langle (\Delta u^{\rho})^2\rangle_{\text{r}} &\simeq& \frac{q^2}{2\pi^2m^2\eta^2a^4R^2}\sum_{n=1}^{p-1}\frac{1}{s_n^2}\nonumber\\
&\simeq&\frac{q^2}{6\pi^2m^2\eta^2a^4R^2}(p^2-1).
\label{limfinal}
\end{eqnarray}
One should note that by analytic continuation, Eq. \eqref{limfinal} describes the approximation $R\ll1$ not only for integer values of $p$ but for any value of it. 

As a matter of fact, it is also possible to get, considering integers values of $p$, an analytic expression for the case where $R\gg 1$. Considering the approximation \eqref{lim1}, we obtain 
\begin{eqnarray}
\langle (\Delta u^{\rho})^2\rangle_{\text{r}} &\simeq& \frac{q^2}{2\pi^2m^2\eta^2a^4R^4}\sum_{n=1}^{p-1}\left[\frac{1}{s_{n}^4}+\frac{2}{s_{n}^2}\right]\nonumber\\
&\simeq&\frac{q^2}{90\pi^2m^2\eta^2a^4R^4}(p^2-1)(p^2+41).
\label{limfinal1.1}
\end{eqnarray}
Again, the expression above is an analytic function of $p$, consequently it is also valid for all values of $p$.

We turn now to the $\phi$-component of the MSD of the particle velocity. Its renormalized expression can obtained using Eqs. \eqref{wf3} and \eqref{Mfun2}. The result is found to be
\begin{eqnarray}
\langle (\Delta u^{\phi})^2\rangle_{\text{r}} &=& \frac{q^2}{\pi^2m^2\eta^2\rho^2a^4}\left[\sideset{}{'}\sum_{n=1}^{[p/2]}f^{(2)}_n(R) -2\int_0^{\infty}d\xi g(\xi,p)f^{(2)}_{\xi}(R) \right],
\label{rphiC2}
\end{eqnarray}
where
\begin{equation}
f^{(2)}_{\gamma}(R) = -\frac{(s_{\gamma}^2-1)}{(1 - R^2s_{\gamma}^2)R^2s_{\gamma}^2} + \frac{(s_{\gamma}^2 - 2)}{4R^3s_{\gamma}^3}\ln\left(\frac{Rs_{\gamma}+1}{Rs_{\gamma}-1}\right)^2.
\label{2comp}
\end{equation}

The limiting case $R\gg 1$ for this component is found to be 
\begin{equation}
f^{(2)}_{\gamma}(R) \simeq\frac{1}{R^4} \left[-\frac{3}{s_{\gamma}^4} + \frac{2}{s_{\gamma}^2}\right],
\label{2compLC}
\end{equation}
which is valid for both terms of the expression \eqref{rphiC2}. However, in the limit $R\ll 1$, we will find the same problem as discussed before in the case of the radial component. Again, to circumvent this, we consider only integer values of $p$ so that the only contribution for the renormalized MSD of the particle velocity comes from the first term in the r.h.s of \eqref{rphiC2}. Thus, the limiting case $R\ll 1$ is obtained as
\begin{eqnarray}
\langle (\Delta u^{\phi})^2\rangle_{\text{r}} &\simeq& -\frac{q^2}{2\pi^2m^2\eta^2\rho^2a^4R^2}\sum_{n=1}^{p-1}\frac{1}{s_n^2}\nonumber\\
&\simeq& -\frac{q^2}{6\pi^2m^2\eta^2\rho^2a^4R^2}(p^2-1).
\label{limfinal2}
\end{eqnarray}
This is an analytic expression in $p$, consequently, valid for all values of $p$ within the limit considered. 

As before, we can also obtain an analytic expression for the approximation $R\gg1$ considering \eqref{rphiC2} only for integer values of $p$. In this case, there is only the first term on the r.h.s of \eqref{rphiC2}, so that the approximated expression \eqref{2compLC} can be used to obtain 
\begin{eqnarray}
\langle (\Delta u^{\phi})^2\rangle_{\text{r}} &\simeq& \frac{q^2}{2\pi^2m^2\eta^2\rho^2a^4R^4}\sum_{n=1}^{p-1}\left[-\frac{3}{s_{n}^4} + \frac{2}{s_{n}^2}\right]\nonumber\\
&\simeq& \frac{q^2}{30\pi^2m^2\eta^2\rho^2a^4R^4}(1-p^4).
\label{limfinal2.1}
\end{eqnarray}
This expression is valid for all values of $p$.

Finally, the $z$-component can be obtained directly by using the Wightman function $W^{(+)}(x,x')$ in \eqref{A4}. It renormalized expression is
\begin{eqnarray}
\langle (\Delta u^z)^2\rangle_{\text{r}} &=& \frac{q^2}{\pi^2m^2\eta^2a^4}\left[\sideset{}{'}\sum_{n=1}^{[p/2]}f^{(3)}_n(R) -2\int_0^{\infty}d\xi g(\xi,p)f^{(3)}_{\xi}(R) \right],
\label{3comp}
\end{eqnarray}
where
\begin{equation}
f^{(3)}_{\gamma}(R) =  \frac{1}{4R^3s_{\gamma}^3}\ln\left(\frac{Rs_{\gamma}+1}{Rs_{\gamma}-1}\right)^2.
\label{sec2eq7}
\end{equation}

The limiting case $R\gg 1$ is obtained as
\begin{equation}
f^{(3)}_{\gamma}(R) \simeq \frac{1}{R^4s_{\gamma}^4},
\label{3compLC}
\end{equation}
which is valid for both terms of \eqref{3comp}.

The limiting case $R\ll1$ can be dealt with as before, considering only integer values of $p$. In this case, only the first term on the r.h.s of Eq. \eqref{3comp} is present. The asymptotic expression is then found to be
\begin{eqnarray}
\langle (\Delta u^{z})^2\rangle_{\text{r}} &\simeq& \frac{q^2}{2\pi^2m^2\eta^2a^4R^2}\sum_{n=1}^{p-1}\frac{1}{s_n^2}\nonumber\\
&\simeq& \frac{q^2}{6\pi^2m^2\eta^2a^4R^2}(p^2-1).
\label{limfinal3}
\end{eqnarray}
This expression, of course, by analytic continuation, is valid for any value of $p$.

For integer values of $p$, taking the expression \eqref{3compLC} for the limiting case $R\gg1$, we find 
\begin{eqnarray}
\langle (\Delta u^{z})^2\rangle_{\text{r}} &\simeq& \frac{q^2}{2\pi^2m^2\eta^2a^4R^4}\sum_{n=1}^{p-1}\frac{1}{s_{n}^4}\nonumber\\
&\simeq& \frac{q^2}{90\pi^2m^2\eta^2a^4R^4}(p^2-1)(p^2+11),
\label{limfinal3.1}
\end{eqnarray}
which is valid for all values of $p$. We can see that all components of the renormalized MSD of the particle velocity have the same dependence at the limiting cases considered. That is, for $R\ll1$ all components behaves as $\frac{1}{R^2}$ while for $R\gg 1$ as $\frac{1}{R^4}$.

We can express all the components of the renormalized MSD of the particle velocity obtained above in terms of the physical velocity which can be written as $v^{\rho} = a u^{\rho}$, $v^{\phi} = a\rho u^{\phi}, v^{z} = a u^{z}$. Therefore, in a compact form we have
\begin{eqnarray}
\langle (\Delta v^{i})^2\rangle_{\text{r}} &=& \frac{q^2\hbar}{\pi^2m^2\eta^2c^3a^2}\left[\sideset{}{'}\sum_{n=1}^{[p/2]}f^{(i)}_n(R) - 2\int_0^{\infty}d\xi g(\xi,p)f^{(i)}_{\xi}(R) \right],
\label{final}
\end{eqnarray}
where the functions $f^{(i)}_{\gamma}(R)$, with $i=1,2,3$, are given by Eqs. \eqref{fun1}, \eqref{2comp} and \eqref{sec2eq7}. Note also that, in order to express the result in the International System of Units we have recovered all constants that were missing. Also, in this case, $R=\frac{2\rho}{c\eta}$. 

Let us now discuss some important features about the above components of the renormalized MSD of the particle velocity. In all three cases, that is, Eqs. \eqref{radialC}, \eqref{rphiC2} e \eqref{3comp}, we have a divergency in $R=0$ which is characteristic of the cosmic string spacetime. Moreover, we also have divergencies occurring every time $Rs_{\gamma}=1$. For the second term on the r.h.s of Eqs. \eqref{radialC}, \eqref{rphiC2} e \eqref{3comp}, $s_{\xi}=\cosh(\xi/2)$ which has its least value occuring for $\xi = 0$ in the integral. Thereby, bigger values of  $s_{\xi}$ makes the equality $Rs_{\gamma}=1$ be satisfied only for $R\leq 1$. All the divergencies in this region are integrable, remaining afterwords only a divergency at $R=1$. As the integral in $\xi$ is not solvable analytically this is hard to see. But we have performed a full analytic approximation and numerical analysis to make sure this is correct.

\begin{figure}[!htb]
\begin{center}
\includegraphics[width=0.4\textwidth]{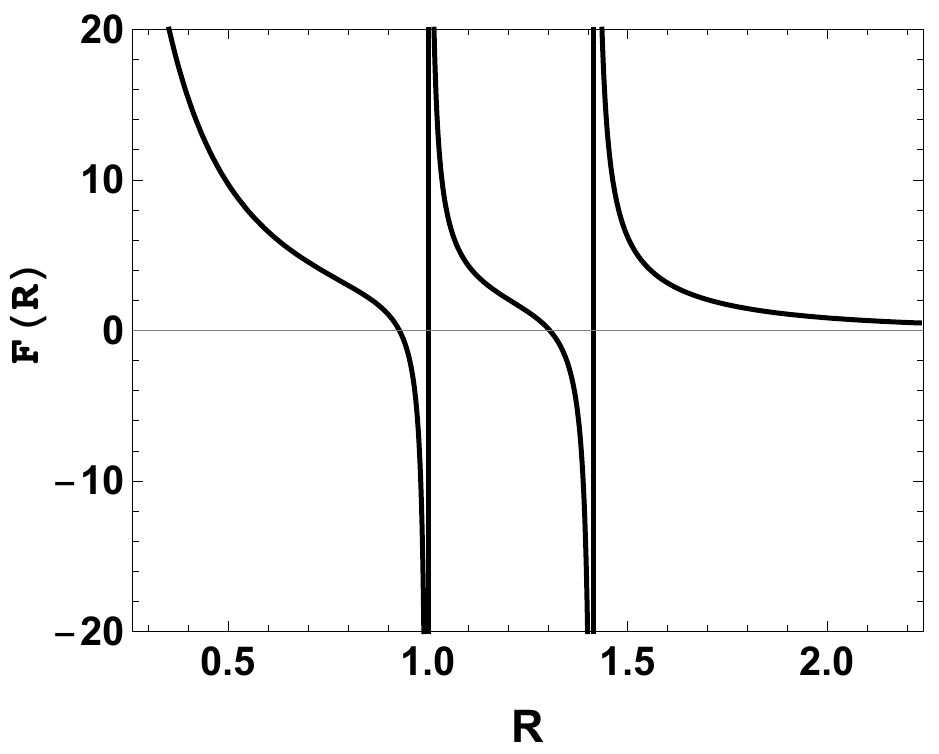}
%\hspace{0.3cm}
\caption{\small{Plot of the dimensionless $\rho$-component of the renormalized MSD of the particle velocity $F(R) = \frac{\pi^2m^2\eta^2c^3a^2\langle (\Delta v^{\rho})^2\rangle_{\text{r}} }{q^2\hbar}$} as a function of the dimensionless radial distance $R=\frac{2\rho}{c\eta}$, considering $p=4$.}
%whilst the one on the right is for $\xi=\frac{1}{6}$. Also, the smaller plot on the right indicates that the curves do not go to zero as they seem to do in the larger plot.
\label{f1}
\end{center}
\end{figure}

 The plot in Fig.\ref{f1} shows as the $\rho$-component of the renormalized MSD of the particle velocity,  in units of  $q^2\hbar/\pi^2m^2\eta^2c^3a^2$, varies with the dimensionless parameter $R$, for $p=4$. This component goes to zero as the distance from the cosmic string increases and to infinity as the distance decreases, in accordance with the asymptotic expressions \eqref{limfinal1.1} and \eqref{limfinal}, respectively. For this value of the cosmic string parameter, the divergences $Rs_{n}=1$ occur at $R=1$ and $R\simeq1.4$, as we can see on the plot of Fig.\ref{f1}. 
 %On should note that this divergencies are analogues to the 'round trip' divergencies reported in Refs.  
%

The equality $Rs_{\gamma}=1$ that makes divergencies to appear can be understood as a time interval for a round trip from the particle location at $\rho s_{\gamma}$ to the cosmic string. This can be seen if we rewrite the equality in terms of the physical variables, i.e., $c\eta = 2\rho s_{\gamma}$. This is similar to the one found in Refs. \cite{Yu:2004gs, Yu:2004tu, Bessa:2017iyk}.  For instance, taking $p=4$ as shown in Fig.\ref{f1} there exists two round trip divergencies. For $n=1$ it starts from the position $\sqrt{2}\rho$, goes until the cosmic string and returns, while for $n=2$ it starts from position $\rho$ goes until the cosmic string and returns to the position $\rho$. The term for $n=3$ provides the same divergency as for $n=1$.
%
%%%%%%%%%%%%%%%%%%%%%%%%%%%%%%%%%
\subsection{Flat boundary}
\label{sec3.2}
%%%%%%%%%%%%%%%%%%%%%%%%%%%%%%%%%
%
Let us now analyse the influence of a flat boundary contribution, orthogonal to the cosmic string, in the MSD of the particle velocity. This flat boundary is implemented by making use of the Dirichlet boundary condition on the massless quantum scalar field. Thereby, under Dirichelet boundary condition the renormalized MSD of the particle velocity can be separated into two contribution: one purely due to the cosmic string topology itself and a second contribution due to the flat boundary, that is,
\begin{eqnarray}
\langle (\Delta u^{i})^2\rangle_{\text{r}} =  \langle (\Delta u^{i})^2\rangle_{\text{r}}^{(\text{p})} + \langle (\Delta u^{i})^2\rangle^{(\text{b,p})},
\label{contr}
\end{eqnarray}
where the Wightman function \eqref{wftotal} has been used. Note that, had Neumann boundary condition been used we would have gotten only a change of sign for the second contribution in the expression above. Note also that, the divergence arising when the coincidence limit, $x'\rightarrow x$, is taken comes from the cosmic string contribution and needs to be renormalized as indicated in the first term on the r.h.s of Eq. \eqref{contr}. This contribution has already been studied in the previous sub-section. Here we want to focus our attention only on the contribution by the boundary, given by the second term on the r.h.s of Eq. \eqref{contr}. The latter can be calculated by make use of the Wightman function $W^{(-)}(x,x') $ in \eqref{A4}. Thereby, after proceeding similarly as in the case without boundary, the expression \eqref{sec2eq3} provides, for the $\rho$-component, the result below:
\begin{eqnarray}
\langle (\Delta u^{\rho})^2\rangle^{(\text{b,p})}&=& \frac{q^2}{\pi^2m^2\eta^2a^4}\left[\sideset{}{'}\sum_{n=-[p/2]}^{[p/2]}h^{(1)}_n(R, Z) - 4\int_0^{\infty}d\xi g(\xi,p)h^{(1)}_{\xi}(R,Z) \right],\nonumber\\
\label{radialCwb}
\end{eqnarray}
where we have introduced the dimensionless variable $Z=\frac{2z}{\eta}$. For this case, the functions above are 
\begin{eqnarray}
h^{(1)}_{\gamma}(R,Z) = -\frac{N^{(1)}_{\gamma}(R,Z)}{8(R^2s_{\gamma}^2 + Z^2 - 1)(R^2 s_{\gamma}^2 + Z^2)^{\frac{5}{2}}},
\label{fun1wb}
\end{eqnarray}
and
\begin{eqnarray}
N^{(1)}_{\gamma}(R,Z) &=& 4 R^2 s_{\gamma}^4 \sqrt{R^2 s_{\gamma}^2 + Z^2} + ( R^2 s_{\gamma}^2 + Z^2-1) \nonumber\\
&\times&[R^2 (s_{\gamma}^2 + s_{\gamma}^4) + (1 - 2 s_{\gamma}^2) Z^2]\ln{\left(\frac{\sqrt{R^2 s_{\gamma}^2 + Z^2} + 1}{\sqrt{R^2 s_{\gamma}^2 + Z^2}-1}\right)^2}.
\label{nume}
\end{eqnarray}
The MSD of the particle velocity above is finite for $n=0$, in contrast with the case without boundary. In fact, the term $n=0$ is the only one that survives in the absence of the cosmic string. So we observe that the presence of the a cosmic string really induces a nonzero Brownian motion codified through the expression of the MSD of the particle velocity \eqref{radialCwb}.

The expression in \eqref{radialCwb} for the $\rho$ component, for a fixed $R$, diverges at the boundary, i.e., at $Z=0$. At this value, the divergent contribution comes from the $n=0$ term of the sum in Eq. \eqref{radialCwb}. In fact, this divergence is a consequence of the imposition of the Dirichlet boundary condition on the plate located at $z=0$. The most relevant contribution in this region is thus
\begin{eqnarray}
\langle (\Delta u^{\rho})^2\rangle^{(\text{b,p})}\simeq -\frac{q^2}{2\pi^2m^2\eta^2a^4Z^2}.
\label{radialCwbzero}
\end{eqnarray}

On the other hand, in order to analyze the behaviour of the $\rho$-component of the MSD of the particle velocity, in the opposite limit, i.e., far away from the boundary $Z\rightarrow\infty$, we can adopt the same strategy we adopted in the case without boundary. Let us then consider only integer values of $p$ in Eq. \eqref{radialCwb}. Thus, in the latter, there will be only the first term on the r.h.s. Taking the limit $Z\rightarrow\infty$ we obtain
\begin{eqnarray}
\langle (\Delta u^{\rho})^2\rangle^{(\text{b,p})}&\simeq& -\frac{3q^2R^2p}{16\pi^2m^2\eta^2a^4Z^6}.
\label{radialCwbZinfity}
\end{eqnarray}
One should note, however, that the above approximation is an analytic expression of $p$ and it is valid for all values of it, not only the integer ones.

The exact expression for the $\phi$-component of the MSD of the particle velocity can be obtained by using the general expressions \eqref{sec2eq3} and Eqs. \eqref{wf3}-\eqref{Mfun3} obtained in the Appendix \ref{A}. This provides
\begin{eqnarray}
\langle (\Delta u^{\phi})^2\rangle^{(\text{b,p})} &=& \frac{q^2}{\pi^2m^2\eta^2\rho^2a^4}\left[\sideset{}{'}\sum_{n=-[p/2]}^{[p/2]}h^{(2)}_n(R) -4\int_0^{\infty}d\xi g(\xi,p)h^{(2)}_{\xi}(R) \right],
\label{phiCwb}
\end{eqnarray}
where
\begin{eqnarray}
h^{(2)}_{\gamma}(R,Z) = -\frac{N^{(2)}_{\gamma}(R,Z)}{8(R^2s_{\gamma}^2 + Z^2 - 1)(R^2 s_{\gamma}^2 + Z^2)^{\frac{5}{2}}},
\label{fun2wb}
\end{eqnarray}
and
\begin{eqnarray}
N^{(2)}_{\gamma}(R,Z) &=& 4 R^2 s_{\gamma}^2(s_{\gamma}^2 - 1) \sqrt{R^2 s_{\gamma}^2 + Z^2} + ( R^2 s_{\gamma}^2 + Z^2-1) \nonumber\\
&\times&[R^2s_{\gamma}^2(s_{\gamma}^2 -2)  + (1 - 2 s_{\gamma}^2) Z^2]\ln{\left(\frac{\sqrt{R^2 s_{\gamma}^2 + Z^2} + 1}{\sqrt{R^2 s_{\gamma}^2 + Z^2}-1}\right)^2}.
\label{nume2}
\end{eqnarray}
At the boundary, the expression for the MSD of the particle velocity diverges with the same power as the radial component, that is,
\begin{eqnarray}
\langle (\Delta u^{\phi})^2\rangle^{(\text{b,p})} &\simeq& -\frac{q^2}{2\pi^2m^2\eta^2\rho^2a^4Z^2}.
\label{phiCwbzero}
\end{eqnarray}
This asymptotic expression comes from the $n=0$ term of the summation in Eq. \eqref{phiCwb}. In the opposite limit, that is, $Z\rightarrow\infty$, we will consider as before only integers values of $p$ in Eq. \eqref{phiCwb}. This provides
\begin{eqnarray}
\langle (\Delta u^{\phi})^2\rangle^{(\text{b,p})} &\simeq& \frac{q^2R^2p}{16\pi^2m^2\eta^2\rho^2a^4Z^6}.
\label{phiCwbZinfty}
\end{eqnarray}
This approximated expression is an analytic function of $p$ and, therefore, is valid for any value of $p$.

Let us turn now to the last component of the MSD of the particle velocity, namely, the $z$-component. The exact expression in this case is given by
\begin{eqnarray}
\langle (\Delta u^{z})^2\rangle^{(\text{b,p})} &=& \frac{q^2}{\pi^2m^2\eta^2a^4}\left[\sideset{}{'}\sum_{n=-[p/2]}^{[p/2]}h^{(3)}_n(R, Z) - 4\int_0^{\infty}d\xi g(\xi,p)h^{(3)}_{\xi}(R,Z) \right],\nonumber\\
\label{zCwb}
\end{eqnarray}
where
\begin{eqnarray}
h^{(3)}_{\gamma}(R,Z) = -\frac{N^{(3)}_{\gamma}(R,Z)}{8(R^2s_{\gamma}^2 + Z^2 - 1)(R^2 s_{\gamma}^2 + Z^2)^{\frac{5}{2}}}.
\label{fun3wb}
\end{eqnarray}
and
\begin{eqnarray}
N^{(3)}_{\sigma}(R,Z) &=& 4 Z^2 \sqrt{R^2 s_{\gamma}^2 + Z^2} + ( R^2 s_{\gamma}^2 + Z^2-1) \nonumber\\
&\times&(2Z^2 - R^2s_{\gamma}^2)\ln{\left(\frac{\sqrt{R^2 s_{\gamma}^2 + Z^2} + 1}{\sqrt{R^2 s_{\gamma}^2 + Z^2}-1}\right)^2}.
\label{nume3}
\end{eqnarray}
\begin{figure}[!htb]
\begin{center}
\includegraphics[width=0.4\textwidth]{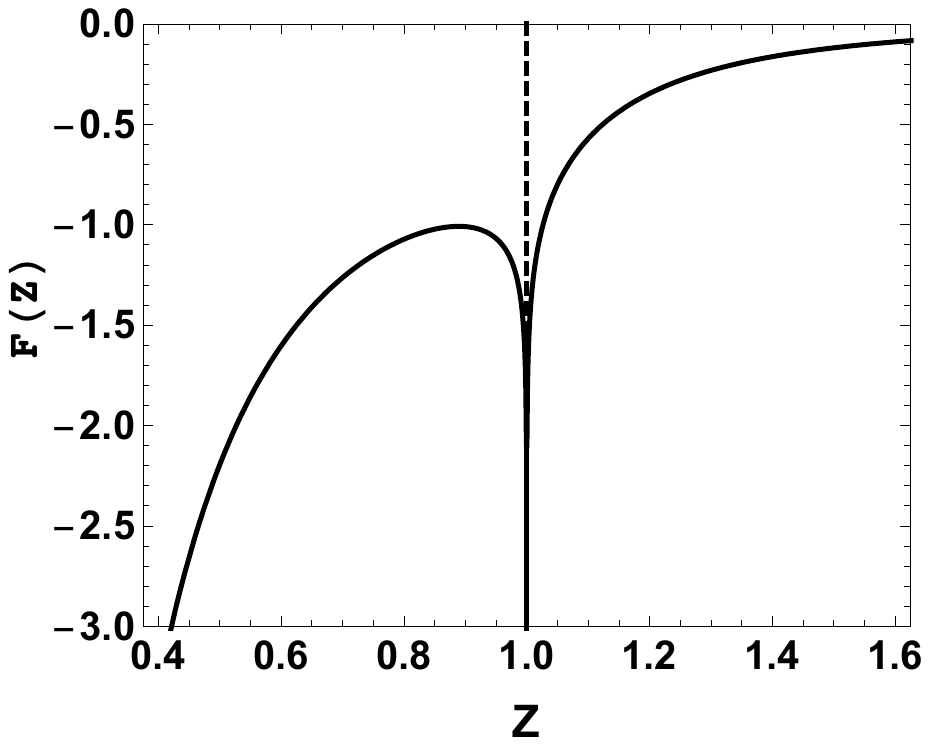}
\hspace{0.3cm}
\includegraphics[width=0.4\textwidth]{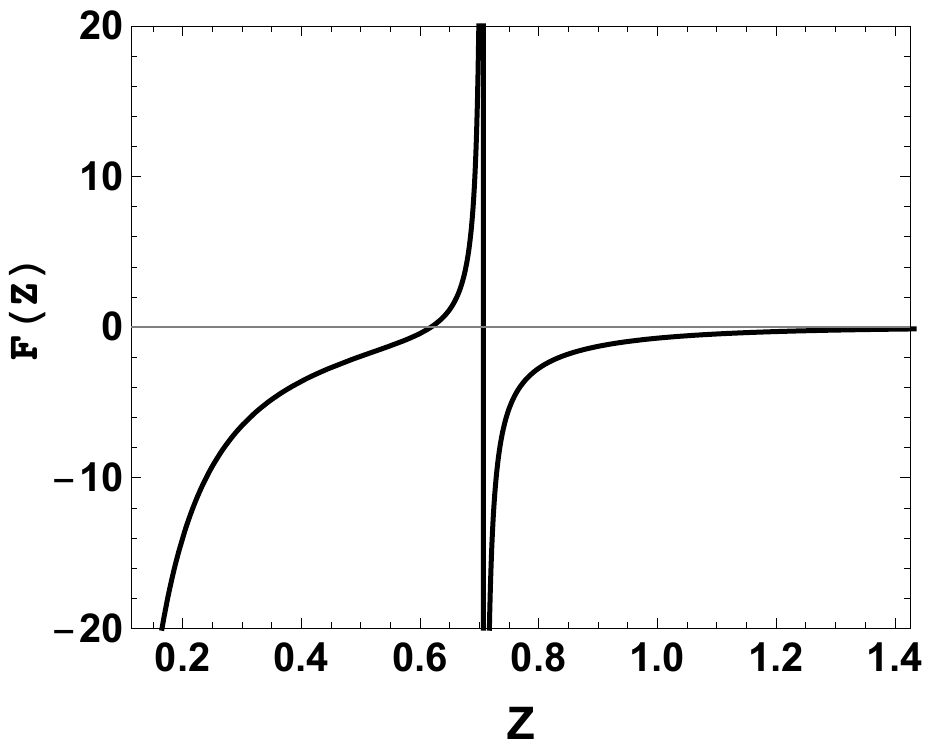}
\caption{\small{Plot of the dimensionless $\rho$-component of the renormalized MSD of the particle velocity $F(Z) = \frac{\pi^2m^2\eta^2c^3a^2\langle (\Delta v^{\rho})^2\rangle_{\text{r}} }{q^2\hbar}$ as a function of the dimensionless distance $Z=\frac{2z}{c\eta}$. The plot on the left is due to the term $n=0$ which does not depend on $R$ and $p$. The plot on the right is for $n\neq 0$, considering $R=1$ and $p=4$.}}
\label{f2}
\end{center}
\end{figure}
Again, the $n=0$ term of the exact expression  \eqref{phiCwb} provides the relevant contribution for the MSD of the particle velocity near to the boundary, i.e.,
\begin{eqnarray}
\langle (\Delta u^{z})^2\rangle^{(\text{b,p})} &\simeq& -\frac{q^2}{2\pi^2m^2\eta^2a^4Z^2}.
\label{zCwb}
\end{eqnarray}
In the opposite limite, $Z\rightarrow\infty$, we can consider again only integer values of $p$. Thereby, we get
\begin{eqnarray}
\langle (\Delta u^{z})^2\rangle^{(\text{b,p})} &\simeq&- \frac{3q^2p}{2\pi^2m^2\eta^2a^4Z^4}.
\label{zCwbZinfty}
\end{eqnarray}
This is an analytic expression of $p$ and, thus, is valid for any value of it. The above expression can be seen as the corresponding one in the Minkowski spacetime, multiplied by the factor $p$. It is worthing mentioning that, for all three components, the MSD of the particle velocity is finite at $R=0$ if $Z\neq 0$.

We can also express all the components of the MSD of the particle velocity induced by the flat boundary in terms of the physical velocity. In a compact form we have
\begin{eqnarray}
\langle (\Delta v^{i})^2\rangle^{(\text{b,p})} &=& \frac{q^2\hbar}{\pi^2m^2\eta^2c^3a^2}\left[\sideset{}{'}\sum_{n=-[p/2]}^{[p/2]}h^{(i)}_n(R,Z) - 4\int_0^{\infty}d\xi g(\xi,p)h^{(i)}_{\xi}(R,Z) \right],
\label{final}
\end{eqnarray}
where the functions $h^{(i)}_{\gamma}(R,Z)$, with $i=1,2,3$, are given by Eqs. \eqref{fun1wb}, \eqref{fun2wb} and \eqref{fun3wb}. Note also that, in order to express the result in International System of Units we have recovered all constants that were missing. Also, in this case, $Z=\frac{2z}{c\eta}$.

A discussion about our results considering the flat boundary can now be made. First of all, the asymptotic behaviours for each component of the renormalized MSD of the particle velocity are similar. Near the flat boundary, the renormalized MSD of the particle velocity expressions goes to infinity as $\frac{1}{Z^2}$ and far away from the boundary goes to zero as $\frac{1}{Z^6}$, except for the $z$-component which behaves as $\frac{1}{Z^4}$. We also have here a divergence comes from the time interval for a round trip from the particle location to the flat boundary. This divergence occurs every time $R^2s_{\gamma}^2 + Z^2 =1$. Or in terms of physical quantities $c\eta = 2\sqrt{\rho s_{\gamma}^2 + z^2}$. We have plotted in Fig.\ref{f2}, on the left side, the renormalized MSD of the particle velocity along the $\rho$ direction,  in units of  $q^2\hbar/\pi^2m^2\eta^2c^3a^2$, for the $n=0$ term, as a function of $Z$ considering $R=1$. This term does not depend of $p$. We can note that there is one round trip divergence at $Z=1$. Out of this point, the renormalized MSD of the particle velocity goes to infinity as the particle approaches the boundary and goes to zero as the particle increases its distance from the boundary, as anticipated by the asymptotic expressions above. 

On the other hand, the plot on the right side of Fig.\ref{f2}, shows the behaviour of the dimensionless MSD of the particle velocity along the $\rho$ direction for $n\neq 0$, as a function of $Z$ considering $R=1$ and $p=4$. For these values of $R$ and $p$ the round trip divergence occurs for $Z=0$ when $n=2$, and for $Z\simeq 0.71$ when $n=1,3$, as showed in the right plot. Depending on the values of $R$ and $p$ theses round trip divergencies can change. As $Z$ increases, we can also note that the renormalized MSD of the particle velocity goes to zero.
\section{conclusions}
\label{con}
We have analyzed the induced Brownian motion of a massive scalar particle as a consequence of quantum vacuum fluctuations of a massless scalar field whose modes propagate in the $(3 + 1)$-dimensional FRW spacetime. In our case, we have also included a linear topological defect known in cosmology as cosmic string. The induced Browninan motion is, thus, due to the FRW geometry as well as the topology of the cosmic string. To analyse the problem we have used the conformal properties of the massless scalar field described in Sec.\ref{sub2.1}. The conformal symmetry connects the FRW spacetime in the presence of a cosmic string with the spacetime only of a cosmic string through the conformal factor $\Omega = a(\eta)$. This made possible for us to use known expressions of the Wightman function in the cosmic string spacetime to calculate the renormalized MSD of the particle velocity using the particle dynamics approach reviewed in Sec.\ref{sub2.2}. The massive particle dynamics is considered in the nonrelativistic regime. 

Besides the geometry and topology of the FRW and cosmic string spacetimes, respectively, we have also considered the effect on the renormalized MSD of the particle velocity of a flat boundary. In this analysis we assumed that the scalar field obeys the Dirichlet boundary condition on a plate. Thus, in Sec.\ref{sec3.1} we have first considered the case with no boundary and obtained exact analytic expression for all component of the renormalized MSD of the particle velocity and showed how they behave near and far away from the cosmic string. Near the linear defect, the renormalized MSD of the particle velocity goes to infinity as $\frac{1}{R^2}$, which is expected as there is a divergence associated to a linear distribution of matter. Far away from the cosmic string, the renormalized MSD of the particle velocity goes to zero as $\frac{1}{R^4}$. Additional divergencies appear when $Rs_{\gamma}=1$ which are associated with the time interval of a round trip for the massive particle to go from its position to the defect and return. These round trip divergencies depend upon the value of the cosmic string parameter $p$. These aspects are exhibited in the plot of Fig.\ref{f1}, considering $p=4$.

Furthermore, we have investigated in Sec.\ref{sec3.2} the influence of a flat boundary orthogonal to the cosmic string on the renormalized MSD of the particle velocity. In this case, it is possible to separate the renormalized MSD of the particle velocity into two contributions: one due to the geometry and topology of the FRW and cosmic string spacetimes and a second one due to the boundary. As we analysed the first contribution in Sec.\ref{sec3.1}, we focus our attention only on the second contribution in Sec.\ref{sec3.2}. We also found exact expressions for this case and showed how all the components behave near and far away from the flat boundary. Near the boundary the renormalized MSD of the particle velocity goes to infinity as $\frac{1}{Z^2}$ while it goes to zero as $\frac{1}{Z^6}$ in the region far from the boundary, except for the $z$-component whose behaviour goes as $\frac{1}{Z^4}$. In the presence of a flat boundary, there are also round trip divergencies whenever $R^2s_{\gamma}^2 + Z^2 = 1$. Depending on the value of $R$, $Z$ and $p$ there can be several round trip divergencies. All these aspects are exhibited in the plots of Fig.\ref{f2} for $p=4$. The left plot is only for $n=0$, which is the case of a nonzero MSD of the particle velocity only due to the flat boundary in the FRW spacetime. The plot on the right is for the contribution due to the renormalized MSD of the particle velocity along the $\rho$ direction when $n\neq 0$, considering $R=1$ and $p=4$.
%
%%%%%%%%%%%%%%%%%%%%%%%%%%%%%%%%%%%%%%%%%%%%%%%%%%%%%%%%%%%%%%%%%%%%
\acknowledgments
H.F.S.M and E.R.B.M are partially supported by CNPq (Conselho Nacional de Desenvolvimento Cient\'ifico e Tecnol\'ogico - Brazil) under grants 305379/2017-8 and 313137/2014-5, respectively.
%%%%%%%%%%%%%%%%%%%%%%%%%%%%%%%%%%%%%%%%%%%%%%%%%%%%%%%%%%%%%%%%%%%%
\appendix
\section{Wightman function}
\label{A}
Let us here derive some important expressions which are necessary for the results obtained in the body of the text. The positive energy Wightman function resulting from the solution \eqref{solution}, obeying Dirichlet boundary condition, can be write in the form \cite{BezerradeMello:2011sm, Mota:2016mhe}
\begin{eqnarray}
W^{(\rm\pm)}(x,x') = \pm \frac{p}{8\pi^2\rho\rho'}\int_{0}^{\infty}dwe^{-\frac{\zeta_{(\mp)}}{2\rho\rho'}w}\sum_{n=-\infty}^{\infty}e^{inp\Delta\phi}I_{p|n|}(w)
\label{wf}
\end{eqnarray}	
where the `plus' sign in $W^{(\rm\pm)}(x,x')$ indicates the Wightman function due to only the cosmic string spacetime and the `minus' sign indicates the Wightman function due to only a flat boundary \cite{BezerradeMello:2011sm} and 
\begin{eqnarray}
\zeta_{(\mp)} = - \Delta \eta^2 + (z\mp z')^2 + \rho^2 + \rho'^2 .
\label{STI}
\end{eqnarray}	
Therefore, the total renormalized Wightman function associated with a massless scalar field in the cosmic string spacetime obeying the Dirichlet boundary condition on the flat boundary is given by
\begin{eqnarray}
W_{\text{r}}(x,x') = W_{\text{r}}^{(+)}(x,x') + W^{(-)}(x,x'),
\label{wftotal}
\end{eqnarray}
where $W_{\text{r}}^{(+)}(x,x') $ is the renormalized Wightman function purely due to the cosmic string topology, i.e., the Wightman function with the Minkowski spacetime divergent contribution already subtracted. Also, $W^{(-)}(x,x')$ is the Wightman function induced by the plain boundary. Note that the Neumann boundary condition would only change the plus sign in Eq. \eqref{wftotal} by the minus sign.

Moreover, in Refs. \cite{deMello:2014ksa, Mota:2016mhe} a summation formula 	to perform the sum in $n$ present in Eq. \eqref{wf} was derived and is given by
\begin{eqnarray}
\sum_{n=-\infty}^{\infty}e^{ipn\Delta\phi}I_{p|n|}(w) &=& \frac{e^w}{p} + \frac{2}{p}\sideset{}{'}\sum_{n=1}^{[q/2]}e^{w\cos\left(\frac{2\pi n}{p}-\Delta\phi\right)}\nonumber\\
&-& \frac{1}{2\pi}\sum_{j=+,-}\int_{0}^{\infty}d\xi\frac{\sin\left[p\left(j\Delta\phi + \pi\right)\right]e^{-w\cosh(\xi)}}{[\cosh(p\xi) - \cos(jp\Delta\phi + p\pi)]},
\label{SF}
\end{eqnarray}
where $[p/2]$ represents the integer part of $p/2$, and the prime on the sign of summation means that in the case $p$ is an integer number the term $n=p/2$ should be taken with the coefficient $1/2$. 
Note that, if $p<2$ the summation contribution should be omitted. Thereby, by using \eqref{SF} the Wightman function is obtained as \cite{BezerradeMello:2011sm, Mota:2016mhe}

\begin{eqnarray}
W^{(\rm \pm)}(x,x')&=&\pm\frac{1}{4\pi^2}\frac{1}{\sigma_0^{(\mp)}} \pm \frac{1}{2\pi^2}\sideset{}{'}\sum_{n=1}^{[p/2]}\frac{1}{\sigma_n^{(\mp)}}\nonumber\\
&\mp&\frac{q}{8\pi^3}\sum_{j=+,-}\int_{0}^{\infty}d\xi\frac{\sin\left[p\left(j\Delta\phi + \pi\right)\right]}{[\cosh(p\xi) - \cos(jp\Delta\phi + p\pi)]}\frac{1}{\sigma_\xi^{(\mp)}},
\label{A4}
\end{eqnarray}
where
\begin{eqnarray}
\sigma_0^{(\mp)} &=& \zeta_{(\mp)} - 2\rho\rho'\cos(\Delta\phi),\nonumber\\
\sigma_n^{(\mp)} &=& \zeta_{(\mp)} - 2\rho\rho'\cos\left(\frac{2\pi n}{p}-\Delta\phi\right),\nonumber\\
\sigma_{\xi}^{(\mp)} &=& \zeta_{(\mp)} + 2\rho\rho'\cosh(\xi).
\label{sec2eq5}
\end{eqnarray}

Another important relation to calculate the azimuthal component of the dispersion in the particle velocity is obtained by taking the derivatives in $\phi$ and $\phi'$ of Eq. \eqref{wf}, that is,

\begin{eqnarray}
\partial_{\phi}\partial_{\phi'}W^{(\rm\pm)}(x,x') = \pm \frac{p^3}{8\pi^2\rho\rho'}\int_{0}^{\infty}dwe^{-\frac{\zeta_{(\mp)}}{2\rho\rho'}w}\sum_{n=-\infty}^{\infty}n^2I_{p|n|}(w).
\label{wf2}
\end{eqnarray}	
Note that, after taking the derivatives in the azimuthal coordinates, we have taken the limit $\phi'\rightarrow\phi$. Next, we can make use of the modified Bessel differential equation

\begin{eqnarray}
\sum_{n=-\infty}^{\infty}n^2I_{p|n|}(w) = \frac{1}{p^2}\left(w^2\partial_w^2 + w\partial_w - w^2\right)\sum_{n=-\infty}^{\infty}I_{p|n|}(w),
\label{relation}
\end{eqnarray}	
to further work out the expression \eqref{wf2}. Thus, 
\begin{eqnarray}
\partial_{\phi}\partial_{\phi'}W^{(\rm\pm)}(x,x') = \pm \frac{p}{8\pi^2\rho\rho'}\int_{0}^{\infty}dwe^{-\frac{\zeta_{(\mp)}}{2\rho\rho'}w}\left(w^2\partial_w^2 + w\partial_w - w^2\right)\sum_{n=-\infty}^{\infty}I_{p|n|}(w).\nonumber\\
\label{wf22}
\end{eqnarray}	
We can again use the summation formula \eqref{SF} for $\Delta\phi =0$ to get
\begin{eqnarray}
\partial_{\phi}\partial_{\phi'}W^{(\rm\pm)}(x,x') = \pm \frac{\rho\rho'}{2\pi^2}\left[\sideset{}{'}\sum_{n=-[q/2]}^{[p/2]}M_n^{(\mp)}(x,x') - \frac{p}{\pi}\int_{0}^{\infty}d\xi\frac{\sin(p\pi)M_{\xi}^{(\mp)}(x,x')}{[\cosh(p\xi) - \cos(p\pi)]}\right],\nonumber\\
\label{wf3}
\end{eqnarray}	
where
\begin{eqnarray}
M_{\gamma}^{(\mp)}(x,x') = 16s_{\gamma}^2(s_{\gamma}^2 - 1)\frac{\rho\rho'}{[\sigma_{\gamma}^{(\mp)}]^3} + (1 - 2s_{\gamma}^2)\frac{1}{[\sigma_{\gamma}^{(\mp)}]^2}.
\label{Mfun}
\end{eqnarray}	
Note that$\gamma$ stands for $n$ in the first term and $\xi$ in the second term of Eq. \eqref{wf3} and $s_n=\sin(n\pi/p)$ and $s_{\xi}=\cosh(\xi/2)$.

We want now to consider the integral in $\eta$ of the functions $M_{\gamma}^{(\mp)}(x,x')$ necessary for the calculation of the $\phi$ component of the velocity dispersion, i.e.,
\begin{eqnarray}
\int_{0}^{\eta}d\eta_1\int_{0}^{\eta}d\eta_2M_{\gamma}^{(+)}(x,x) = \frac{1}{\eta^2c^4}\left[-\frac{(s_{\gamma}^2 - 1)}{R^2s_{\gamma}^2(1-R^2s_{\gamma}^2)} + \frac{(s_{\gamma}^2 - 2)}{4R^3s_{\gamma}^3}\ln\left(\frac{Rs_{\gamma} + 1}{Rs_{\gamma} - 1}\right)^2\right]
\label{Mfun2}
\end{eqnarray}	
and
\begin{eqnarray}
\int_{0}^{\eta}d\eta_1\int_{0}^{\eta}d\eta_2M_{\gamma}^{(-)}(x,x) &=& \frac{1}{4\eta^2c^4(R^2s_{\gamma}^2+Z^2)^{\frac{5}{2}}}\left[\frac{4R^2s_{\gamma}^2(s_{\gamma}^2-1)\sqrt{R^2s_{\gamma}^2 +Z^2}}{(R^2s_{\gamma}^2 + Z^2 -1)}\right.\nonumber\\
&&\left. + [(1-2s_{\gamma}^2)Z^2 + (s_{\gamma}^2-2)R^2s_{\gamma}^2]\ln\left(\frac{\sqrt{R^2s_{\gamma}^2 +Z^2} + 1}{\sqrt{R^2s_{\gamma}^2 +Z^2} - 1}\right)^2\right].\nonumber\\
\label{Mfun3}
\end{eqnarray}	
Therefore, the expressions \eqref{wf3}-\eqref{Mfun3} are all necessary to calculate Eqs. \eqref{rphiC2} and \eqref{phiCwb}.
%
	
%\bibliography{refs}
\end{document}